\documentclass[journal]{IEEEtran}
\usepackage{amsmath,amsfonts,amssymb,mathrsfs}
\usepackage{graphicx,xcolor,soul}
\usepackage{cite}
\usepackage{multirow}
\usepackage{tabularx}
\DeclareMathOperator{\sinc}{sinc}
\DeclareMathOperator{\sincsq}{sq}

\begin{document}
%
\title{Reconfigurable and Real-Time Nyquist OTDM Demultiplexing in Silicon Photonics}
%
%
%

\author{Arijit~Misra,~
        Karanveer~Singh,~Janosch~Meier,~Christian~Kress,~Tobias~Schwabe,~Stefan~Preu{\ss}ler,~J.~Christoph~Scheytt,~
        and~Thomas~Schneider
\thanks{This work was supported in part by the Deutsche Forschungsgemeinschaft (DFG, German Research Foundation) under grant numbers - 322402243, 403154102, 424608109, 424608271, 424607946, 424608191, and in part by the German Federal Ministry of Education and Research (BMBF) under funding code 13N14879.}
\thanks{A. Misra, K. Singh, J. Meier, S. Preu{\ss}ler and T. Schneider are with the THz Photonics Group, Technische Universit\"at Braunschweig, Schleinitzstra{\ss}e 22, 38106 Braunschweig, Germany (Coressponding author: Arijit Misra, e-mail: arijit.misra@ihf.tu-bs.de).}
\thanks{C. Kress, T. Schwabe, and J. C. Scheytt are with the Paderborn University, System and Circuit Technology, Heinz Nixdorf Institute, F\"urstenalle 11, 33098, Paderborn, Germany.}}
\maketitle

\begin{abstract} 
We demonstrate for the first time, to the best of our knowledge, reconfigurable and real-time orthogonal time-domain demultiplexing of coherent multilevel Nyquist signals in silicon photonics. No external pulse source is needed and frequency-time coherence is used to sample the incoming Nyquist OTDM signal with orthogonal sinc-shaped Nyquist pulse sequences using Mach-Zehnder modulators. All the parameters such as bandwidth and channel selection are completely tunable in the electrical domain. The feasibility of this scheme is demonstrated through a demultiplexing experiment over the entire C-band (1530 nm - 1550 nm), employing 24 Gbaud Nyquist QAM signals due to experimental constraints on the transmitter side. However, the silicon Mach-Zehnder modulator with a 3-dB bandwidth of only 16 GHz can demultiplex Nyquist pulses of 90 GHz optical bandwidth suggesting a possibility to reach symbol rates up to 90 GBd in an integrated Nyquist transceiver.  
\end{abstract}


\IEEEpeerreviewmaketitle

\section{Introduction}
\label{sec:intro} 
\IEEEPARstart{S}{urging} data traffic in high-speed communication, the internet-of-things and datacenters demand a further scaling of multiplexing and transmission techniques, along with upgraded transceiver designs capable of handling higher modulation formats adaptively and flexibly on a power efficient, small footprint, and economical platform like silicon photonics. Along with the efforts in parallel transmission using multiplexing in space, wavelength and polarization the enhancement of single carrier line rates is also essential in this regard \cite{Hu2018,Lavery2019}. Time-domain multiplexing (TDM) facilitates single carrier line rate enhancement in a wavelength division multiplexing (WDM) system without increasing the optoelectronic bandwidth of available transceiver components \cite{ITU}.\par
Sinc-shaped Nyquist pulse based orthogonal-TDM (OTDM) offers the most efficient way of time-domain multiplexing owing to the zero inter-symbol-interference (ISI) between such pulses, low peak-to-average power ratio (PAPR), and tolerance to linear and nonlinear impairments in high data rate channels \cite{Nakazawa2012,Schmogrow2012a,Soto2013,Bosco2010,Hirooka2018,Yoshida2021, LeutholdJuergandBres2015}. This allows several time-domain channels to be multiplexed, if the data is modulated on such pulses and then interleaved with a specific time delay dictated by the orthogonality condition \cite{Nakazawa2012, Soto2013}. Such a sinc-shaped Nyquist pulse
sequence based transceiver is signal-format-transparent with the capability of transmitting versatile data formats on a single carrier \cite{Misra2021}. Additionally, for a multi-carrier or WDM system, the need for any guardband can be precluded due to the rectangular shape of the singular wavelength channels \cite{Misra2021}. So, a WDM system based on sinc-shaped Nyquist pulses can exploit the available channel bandwidth for data transmission in the maximum possible way. \par 
There has been abundant exploration on sinc-shaped Nyquist pulse synthesis and multiplexing techniques over the years. However, the reception techniques are limited to a few. Demultiplexing of sinc-shaped orthogonal time-domain multiplexed channels requires either the sampling with much higher sampling rates and narrower time gates, or a sinc-shaped orthogonal time-domain gate that extracts the concerned time-domain channel without interference. The first method is realized by very high-bandwidth digital signal processing (DSP) \cite{DaSilva2016}, or by polarization-sensitive nonlinear optical sampling of synchronous control pulse trains of much higher bandwidth in a nonlinear optical loop mirror (NOLM) \cite{Mulvad2010, Nakazawa2012, Hirooka2015, Hirooka2018, Yoshida2021}. For the sinc-shaped orthogonal time-domain gate, linear optical sampling in a coherent receiver using Nyquist pulse sequences as the local oscillator (LO) \cite{Harako2014a,Tan2015,Tan2016}, optical Fourier transform (OFT) \cite{Hu2014}, and optical parametric amplification \cite{Hansryd2001} have been utilized. However, these receiver configurations suffer from significant drawbacks such as complex system architectures for nonlinear optical interactions, static configuration due to scarcity of reconfigurable external sampling pulse sources in terms of central wavelength and pulse width, need for sampling pulses with bandwidth much higher than the signal bandwidth, or intensive electronics and computational hardware requirements in terms of bandwidth for the DSP.\par
In contrast, time-frequency coherence based all-optical sinc-shaped pulse sequence sampling by a Mach-Zehnder Modulator (MZM) provides a much easier solution with the existing state-of-the-art components \cite{Preussler2016, Misra2019OE, Meier2019,Meier2019a, Misra2019b}. The method utilizes spectral convolution by creating equal, phase-locked spectral replicas of the incoming signal to achieve a full-field sampling in the time-domain. It neither requires any reconfigurable Nyquist pulse sources operating at different spectral bands, nor any optical filters or tunable delay lines. Channel selection is achieved merely by phase shifts in the electrical driving signals to the modulator. \par 
Besides simplicity, this method is also agile by virtue of multi-wavelength operation with fully controllable operational bandwidth. A high bandwidth Nyquist OTDM signal can be demultiplexed into $N$ low bandwidth signal channels. The bandwidth of the receiver electronics can be as low as $1/(2N)$ times the overall single carrier symbol rate. Additionally, it is possible to demultiplex higher order quadrature amplitude modulated (QAM) channels in a real-time and dynamic manner with a single MZM \cite{Misra2021}.\par
Based on the convenient setup and the basic integrability of the optical and electronic components, this sampling technique is apt for realization using integrated photonics platforms like silicon photonics, indium phosphite, or lithium niobate on insulator. Many commercial foundries offer silicon MZMs, germanium photodetectors, and other standard photonic devices as part of their process design kits \cite{Siew2021,IHP}. In addition, silicon leverages matured semiconductor processing technology for electronic integration along with photonics. So far, a few non-orthogonal optical time-domain multiplexing or serializer circuits based on Gaussian pulses have been reported in silicon photonics \cite{Aboketaf2010,Verbist2019}. Recently, sinc-shaped Nyquist pulse sequence generation with various silicon based MZM structures have been demonstrated \cite{Liu2020b,Misra2019OE,Misra2020c,Liu2021}, and their quality has been analyzed in detail \cite{De2021}. However, the demonstration of time-domain demultiplexing in any integrated platform is yet to be reported.\par

In this article, we present for the first time, to the best of our knowledge, the experimental demonstration of orthogonal time-domain demultiplexing in silicon photonics. High capacity Nyquist QAM signals have been demultiplexed to three lower capacity channels in real-time by linear sinc-shaped Nyquist pulse sequence sampling using electronic-photonic co-integrated dual-drive silicon MZMs with high modulation efficiency (external $V_\pi$ = 420 mV), low power consumption, and high DC extinction ratio (40 dB). 16QAM and QPSK modulation formats have been received after up to 30 km single-mode fiber transmission at multiple wavelengths over the C-band of optical telecommunication. Investigations on sinc-shaped pulse sequence generation of different bandwidths and repetition rates have also been presented with a demonstrated capability of synthesizing sinc-shaped Nyquist pulse sequences of 90 GHz bandwidth. The experimental results suggest potential silicon coherent transceivers with single carrier symbol rates of 90 GBd. 
\section{MZM based Nyquist time-domain demultiplexing}
\label{sec:principles}

\begin{figure}[!ht]
   \centering
   \includegraphics[width=\columnwidth]{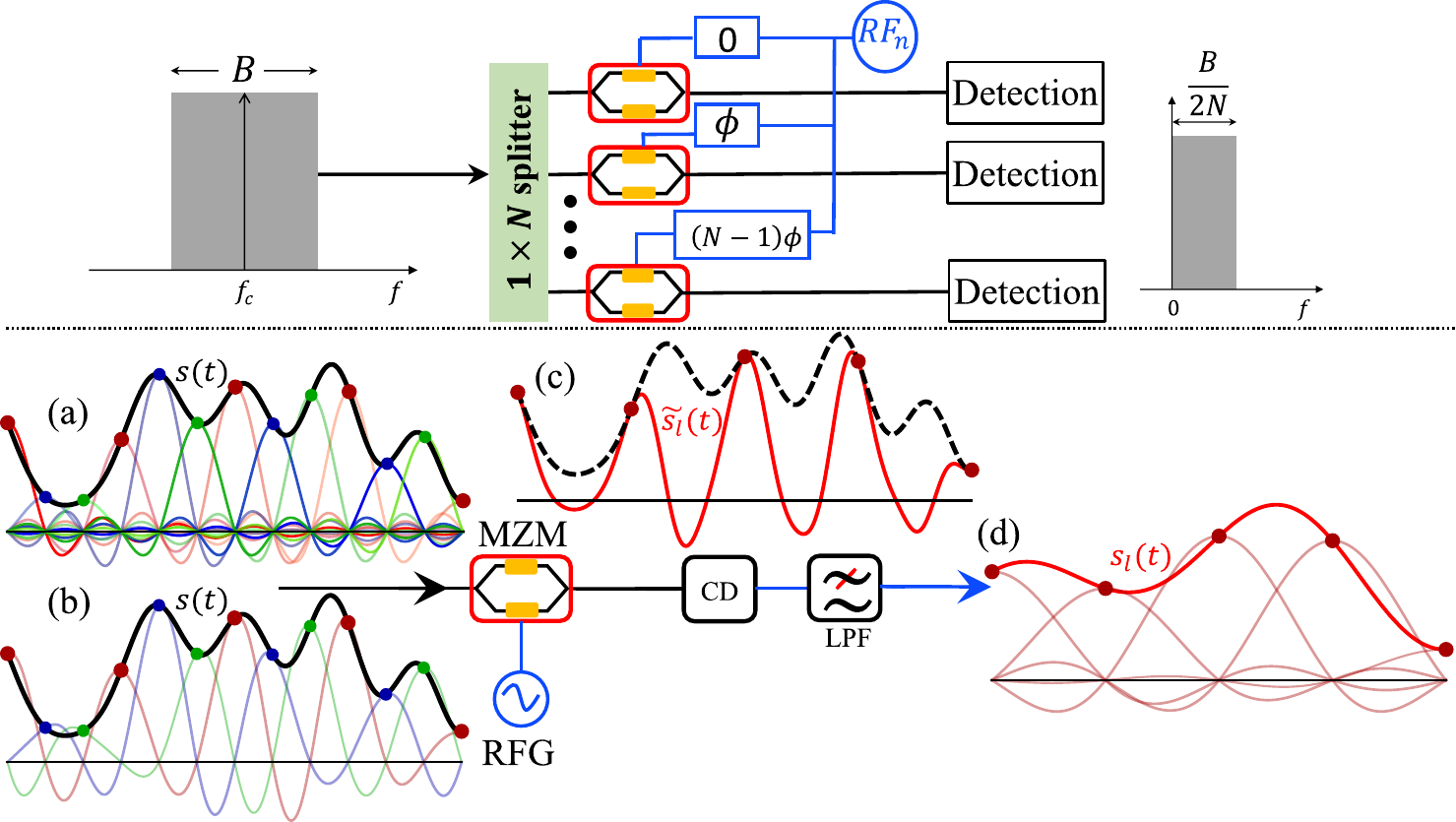}
   \caption{Basic configuration of orthogonal time-domain demultiplexing using MZMs and sinc-shaped Nyquist pulse sequences. In the upper part a complete generalized system is shown, while the lower part presents the signal processing steps for one of the branches, if the MZM is driven with one sinusoidal radio frequency. The black lines correspond to optical and the blue to electrical connections. MZM: Mach-Zehnder modulator, RFG: radio frequency generator, LPF: electrical lowpass filter, CD: coherent detector.}
   \label{fig:concept} 
\end{figure}

The basic principle behind the integrated orthogonal time-domain demultiplexing system is the multiplication between the received modulated Nyquist signal of bandwidth $B$ and an unmodulated sinc-shaped Nyquist pulse sequence with the same bandwidth in each of the $N$ parallel receiver branches. This multiplication reduces the data rate of the incoming signal by the factor $N$. This is shown for $N=3$ for the incoming signal Fig.\ref{fig:concept} (a) and (b) with the red, green and blue dots. By the multiplication with a sinc-pulse sequence with the right time shift, as the red one in Fig.\ref{fig:concept} (c), only the red points (or data signals) will be extracted. In a second and third branch the green and blue points can be extracted at the same time. As result the detector and signal processing electronics in each branch just require a bandwidth of $B/(2N)$. \par
For all of the optical demultiplexing techniques reported so far, external pulse sources are required \cite{Mulvad2010, Nakazawa2012, Hirooka2015, Hirooka2018, Yoshida2021,Harako2014a,Tan2015,Tan2016}. For the method presented here, however, the incoming signal spectrum is convolved with a flat, equispaced, phase-locked optical frequency comb. In the time-domain, this results in the desired multiplication of the signal with a sinc-shaped Nyquist pulse sequence \cite{Preussler2016,Misra2019OE}. The multiplication occurs for
the full-field, thereby achieving coherent sampling \cite{Meier2019,Meier2019a}.\par 
The operation principle for a demultiplexer with $N$ branches is presented in the upper part of Fig. \ref{fig:concept}. A Nyquist signal of optical bandwidth $B$ is demultiplexed in parallel in $N$ odd numbers of branches with MZMs, each driven with $n=(N-1)/2$ equispaced, phase-locked radio frequencies ($RF_n$). In each modulator, the input signal undergoes a sinc-sequence sampling, or a spectral convolution with an $N$-line, flat, rectangular, phase-locked optical frequency comb. Similar to ideal sinc pulses, sinc-sequences are orthogonal. The sampling instances are controlled precisely by the RF phase shifts between the inputs to the modulator. Hence the orthogonality can be controlled entirely in the RF domain. Moreover, the frequency of the highest RF tone and the RF bandwidth of the modulator has to be just $B/3$ for a three branch system. Even a modulator bandwidth of $B/5$ is possible, as we will show in the results section. \par
After the orthogonal time-domain sampling with the correct input phase shifts, the signals in each branch are detected by a detection system with a baseband bandwidth of $B/(2N)$. Thereby, the Nyquist signal with an optical bandwidth $B$ is demultiplexed into $N$ electrical Nyquist signals with a bandwidth of $B/(2N)$. \par 

In the following paragraphs, we present a mathematical description of this process. An optical Nyquist signal can be created by modulating an electrical Nyquist-shaped data signal on a continuous wave (CW) optical carrier \cite{Schmogrow12}. However, it can also be generated all-optically via modulation and time-domain multiplexing of spectrally shaped mode-locked laser pulses \cite{Nakazawa2012} or Nyquist pulse-sequences synthesized by one or more cascaded modulators \cite{Soto2013, SotoCleo, Misra2021}. Therefore, we shall show how regardless of the origin, the high-bandwidth Nyquist channel can be demultiplexed to several low-bandwidth signals using this method. \par
In general, a Nyquist signal can be seen as the infinite sum of time-shifted ideal sinc-pulses weighted with the data symbols. Such a signal $s(t)$ (see Fig. \ref{fig:concept}(a)) will have a rectangular bandwidth of $B$ in the optical domain.
An identical Nyquist signal $s(t)$ can be created as the sum of $N$ time-shifted sinc-shaped pulse sequences, weighted with $N$ data streams of lower symbol rates.  Figure \ref{fig:concept}(b) shows this for a sinc-sequence with two zero-crossings between the peaks, corresponding to a 3-line comb. In the frequency domain, such a sinc-sequence is a flat, three-line frequency comb of bandwidth $B$ \cite{Soto2013, Misra2019OE,Meier2019}. This three-line sinc-sequence can be used to transmit three multiplexed Nyquist channels with a symbol rate of $B/3$ for each channel, and thus an aggregated symbol rate of $B$ in the optical bandwidth $B$. So, regardless if ideal sinc pulses or sinc-sequences are considered, the resultant time-domain signals are identical.\par

If $N$ Nyquist signals of symbol rate $B/N$ are modulated on sinc-sequences of bandwidth $B$ having $N-1$ zero crossings between two consecutive pulse peaks and multiplexed orthogonally in time, then the resultant signal of symbol rate $B$ can be written as,
\begin{equation}
\centering
s(t)=\sum_{l=1}^{N} s_l\left(t\right)\cdot\sincsq_{N,B}\left(t-\frac{l-1}{B}\right),
\label{Eq:1}
\end{equation}
with $l=1,2,...,N$, and $N$ as an odd number.\par
Here, following the Nyquist-Shannon sampling theorem, one of the $N$ different Nyquist data streams with a symbol rate $B/N$ can be expressed as,
\begin{equation}
\centering
s_l(t)=\sum_{k=-\infty}^{\infty} s_l\left(\frac{kN}{B}+\frac{l-1}{B}\right)\cdot\sinc\left(\frac{B}{N}t-\frac{l-1}{N}-k\right)
\label{Eq:2}
\end{equation}
Moreover, the sinc-shaped pulse sequence in Eq. (\ref{Eq:1}) can be defined as the sum of time-shifted ideal sinc-pulses as,
\begin{equation}
\centering
\sincsq_{N,B}(t)=\sum_{k=-\infty}^{\infty} \sinc(Bt-kN).
\label{Eq:3}
\end{equation}
Due to orthogonality of the sinc-sequences, the signal $s(t)$ has an overall symbol rate of $B$ and following the definition in Eq. (\ref{Eq:1}), a baseband bandwidth of $B/2$. Accordingly, it can be identified as a Nyquist signal and therefore, following the Nyquist-Shannon sampling theorem, it can be expressed as the sum of ideal time-shifted sinc-pulses, weighted with the data symbols: 
\begin{equation}
\centering
s(t)=\sum_{k=-\infty}^{\infty}  s\left(\frac{k}{B}\right)\cdot\sinc\left(Bt-k\right).
\label{Eq:4}
\end{equation}
The operation of orthogonal demultiplexing of $s(t)$ into several low bandwidth Nyquist signals can be seen as a two-step process. First the incoming signal is sampled with sinc-sequences and second it is filtered or detected with low bandwidth. The sampled signal in one of the branches as shown by the red curve in Fig. \ref{fig:concept}(c) can be mathematically expressed as,
\begin{equation}
\centering
\tilde{s_l}(t)=s(t)\cdot \sincsq_{N,B}\left(t-\frac{l-1}{B}\right).
\label{Eq:5}
\end{equation}
The sinc-sequence sampled signal can be filtered in the optical domain by an optical bandpass filter of bandwidth $B/N$ to create an optical analog signal for further all-optical signal processing purposes. However, baseband analog or digital filtering can also be done after demodulation with a low-pass filter of bandwidth $B/(2N)$. Alternatively, a detection system of that bandwidth can be used. The output signal after filtering is given by \cite{Misra2021},
\begin{equation}
\begin{aligned}
&\left[\mathscr{F}^{-1}_f \left(\left[\mathscr{F}_t\left(\tilde{s_l}(t)\right)\right](f)\cdot \sqcap_{B/N}(f)\right)\right](t)\\
&=\frac{1}{N}\sum_{k=-\infty}^{\infty} s_l\left(\frac{kN}{B}+\frac{l-1}{B}\right)\cdot\sinc\left(\frac{B}{N}t-\frac{l-1}{N}-k\right)\\
&=\frac{1}{N}s_l(t).
\end{aligned}
\label{Eq:6}
\end{equation}
Here, the low bandwidth detection system or the low pass filter is defined as a rectangular function $\sqcap_{\frac{B}{N}}(f)$ equal to 1 for $\lvert f \rvert < \frac{B}{2N}$, $0.5$ for $\lvert f \rvert = \frac{B}{2N}$, and 0, elsewhere.\par
Therefore, if the incoming signal is split into $N$ branches, the demultiplexing can be carried out in parallel and real-time with MZMs placed in each branch. The sinc-sequences have to be time-shifted by $1/B$ relative to one another. This corresponds to a relative phase shift of $2\pi/N$ between the driving RF signals. In the simplest case, if the MZM is driven with one RF tone, this results in three low-bandwidth signals at the output, corresponding to $l$=1,2, and 3. The red sampling points $\tilde{s_l}(t)$ in Fig. \ref{fig:concept}(b)) were achieved with the red sinc-pulse sequence for sampling as shown in Fig. \ref{fig:concept}(d). The sampling points in between can be achieved with two sinc pulse sequences phase shifted by $120^\circ$ and $240^\circ$, as shown with the blue and green dots in Fig. \ref{fig:concept}(a) and \ref{fig:concept}(b).
\section{Experiment and Results}

In this section, the proof of concept experimental details, characterization of the si-MZM, and data transmission results are presented. First, the performance of the silicon photonic MZM as a Nyquist time-domain demultiplexer has been investigated over the entire C-band (1530 nm - 1560 nm). Later the generation of Nyquist pulses of different pulse widths and repetition rates is investigated to assess the capacity of a single carrier Nyquist transceiver based on the presented MZM. We did not create one high symbol rate Nyquist signal for the experiments. Instead, three low symbol rate signals were multiplexed with sinc-sequences to create one high-capacity Nyquist channel. Afterward, we demultiplexed each of them separately by the silicon photonic MZM. Following the theory section, this is equivalent to the demultiplexing of one single, high data rate Nyquist signal.
\subsection{Experimental setup and device}
\begin{figure}[!t]
   \centering
   \includegraphics[width=\columnwidth]{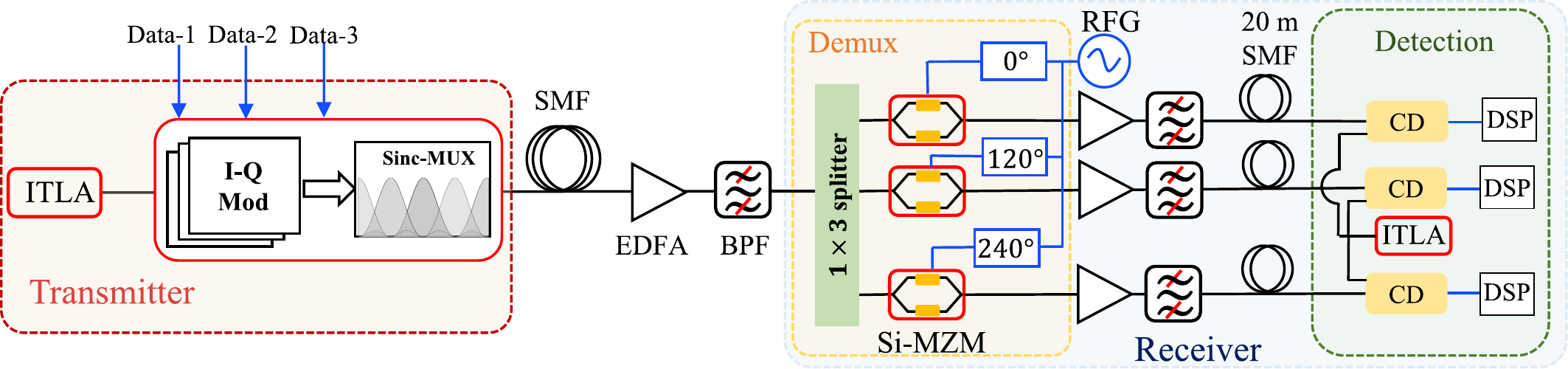}
   \caption{Schematic illustration of the experimental setup. A single carrier from an integrated tunable laser assembly (ITLA) is modulated with three different data channels and then orthogonally multiplexed using sinc-shaped pulse sequences to create a high-capacity Nyquist channel. After transmission through single mode fiber (SMF), the channels are demultiplexed by the Si-MZM before detection by a coherent detector. EDFA: Er-doped fiber amplifier, DSP: digital signal processing. }
   \label{fig:setup} 
\end{figure}
For the experimental demonstration, a setup as illustrated in Fig. \ref{fig:setup} was adopted.  A wavelength-tunable CW laser was modulated with three different data streams and orthogonally multiplexed with sinc-shaped Nyquist pulse sequences in the time-domain. After transmission through standard single mode fiber (SMF), the multiplexed signal was subjected to the silicon electronic-photonic co-integrated Mach-Zehnder modulator. After 20 m SMF transmission, a coherent detector detected the demultiplexed signals with another continuous wave
laser used as the local oscillator. The 20 m SMF was necessary since the integrated device measurement setup was placed in another laboratory room than the asynchronous detector. We used post-detection digital low-pass filtering in the baseband to mimic a low-bandwidth detection system. A coherent modulation analyzer (Tektronix-OM1106) performed the required digital signal processing (DSP) of the recorded waveforms in a real-time oscilloscope (Tektronix DPO73304) for the visualization of symbol constellations and the measurement of other performance metrics like $Q$-factor and error vector magnitude (EVM). Forward error correction, pre-distortion, and nonlinearity compensation were not applied in the experiment. 1\% of the output power after the chip was subjected to an optical spectrum analyzer (OSA, Yokogawa AQ6370) for spectral measurements. The limited sampling rate of the used arbitrary waveform generators for data signal generation prohibited the possibility of reaching higher single carrier optical bandwidths beyond 24 GHz for the sinc-sequence multiplexed channels. As the transmitter was limiting the symbol rate to $B=24$ GBd, the RF input to the modulator was 8 GHz ($B/3$), and the receiver was restricted in the baseband to 4 GHz ($B/6$) via low pass digital filtering. EDFAs along with BPFs were used to amplify the optical signals when required. Polarization diversity was not employed in the proof of concept experiment.\par

\begin{figure}[!ht]
   \centering
   \includegraphics[width=0.85\columnwidth]{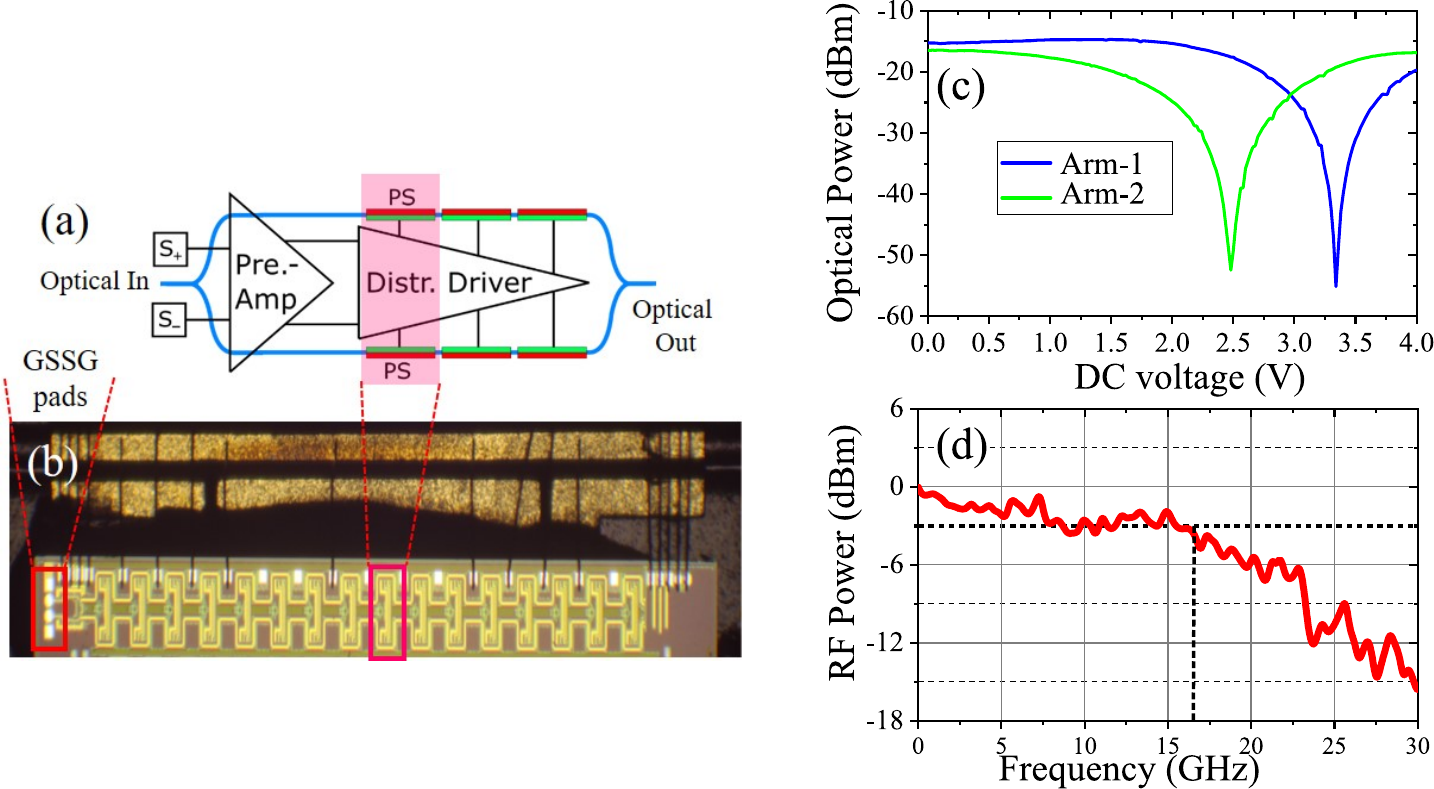}
   \caption{ (a) Block diagram of the fabricated MZM with a pre-amplifier and a distributed driver that applies complementary signal inputs to the segmented phase shifters in the two arms. (b) Chip picture: Bond wires connect chip pads to printed circuit board lines for DC connections. The driving signal to the MZM is applied via a probe in ground-signal-signal-ground (GSSG) configuration. (c) Optical transfer characterization of the two arms in terms of DC input to the heaters, integrated with the two arms of the Si-MZM. (d) Measured electro-optic response of the MZM. The 3 dB bandwidth of 16 GHz is indicated with a dashed line. }
   \label{fig:chip} 
\end{figure}
The Si-MZM chip was fabricated in a BiCMOS technology from IHP (SG25H5\_ePIC). This technology allows monolithic integration of high-frequency driver circuits along with the electro-optical phase shifters for the MZMs to enhance the modulation efficiency and bandwidth of the modulator. The modulator used for the experiment is a segmented MZM with distributed linear drivers, as shown in Fig. \ref{fig:chip}(a) and \ref{fig:chip}(b). The overall length of the device is 6.25 mm. As a metric for modulation efficiency, the measured $V_{\pi}$ of 420 mV is considerably lower than the best-reported state-of-the-art monolithic approaches \cite{Rito2016}. As shown in Fig. \ref{fig:chip}(a), the MZM chip contains segmented MZM and fully differential drive electronics. The on-chip driver incorporates a pre-amplifier with a 100 Ohm differential input resistance. The subsequent main driver is designed in a distributed manner for driving the phase shifter segments (PS) of the MZM using a push-pull scheme. To achieve full phase shift for high-speed operation, and considering the maximum achievable voltage swing from the SiGe transistors limited by collector-emitter breakdown limitations, the total effective phase shifter length for each arm was chosen to be 6.24 mm. For the DC bias point setting of the MZM, additional on-chip heater structures, controlled via external voltage sources, have been placed in each arm. Electrical signals are applied either by bondwires for DC connections (Fig. \ref{fig:chip}(b)) or via probes for the RF signals. For a full $2 \pi$ phase shift of the MZM bias point, maximum electrical power of 90 mW must be applied to the thermal phase shifters. The DC extinction ratio was measured to be 40 dB and 37 dB for the two arms by sweeping the input voltage to the heaters (Fig. \ref{fig:chip}(c)). The electro-optic response of the modulator is shown in Fig. \ref{fig:chip}(d).\par
\subsection{Results and discussion}

\begin{figure}[!ht]
   \centering
   \includegraphics[width=0.85\columnwidth]{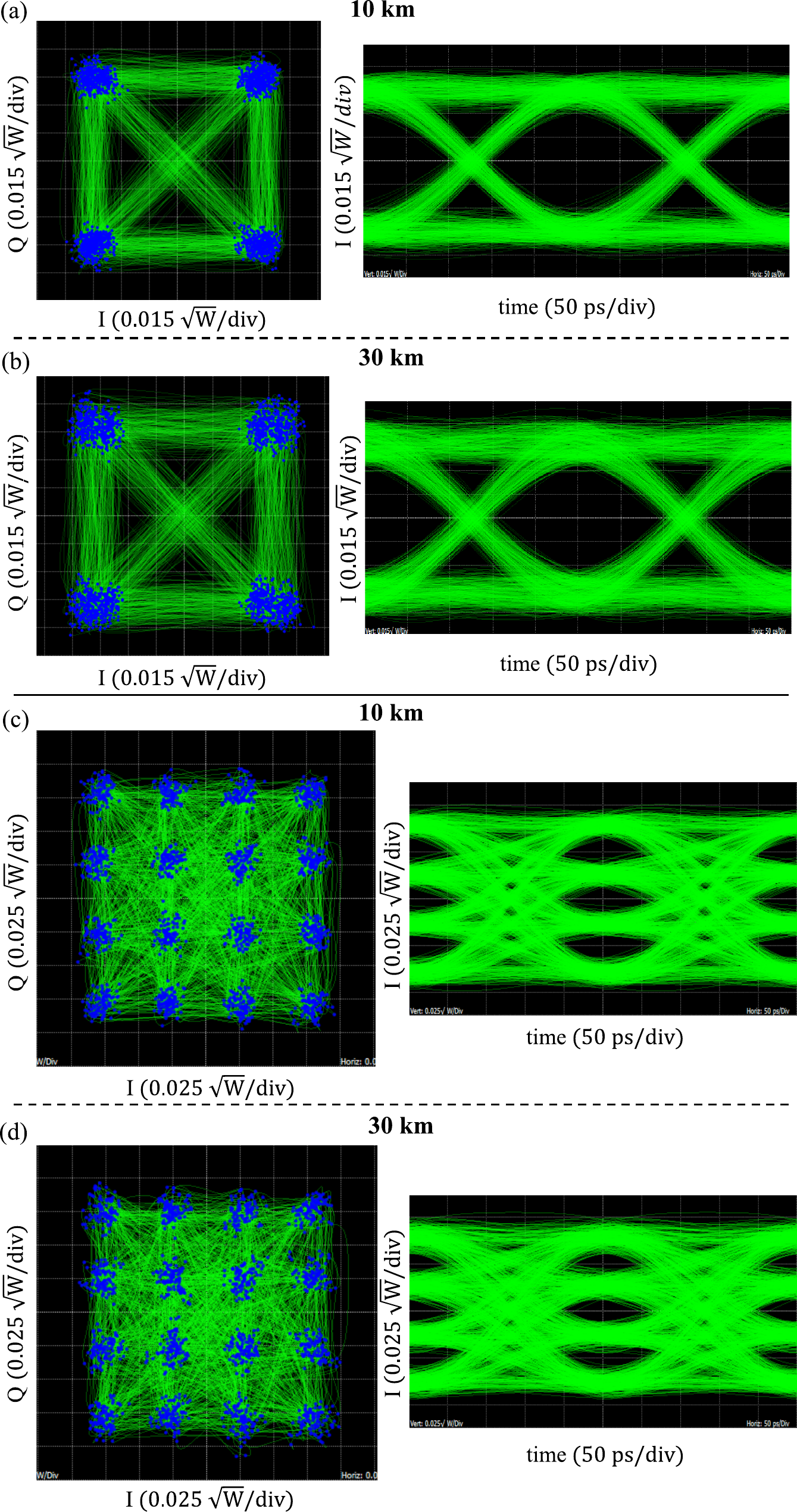}
   \caption{ Measured Symbol constellation and eye diagram for one demultiplexed 4 GBd QPSK (a,b), and a 4 GBd 16-QAM signal (c,d) from an orthogonal sinc-sequence multiplexed channel with a symbol rate of 12 GBd (24 GHz bandwidth) originating from a single carrier at 193.4 THz. Experiments were carried out for 10 km (a,c) and 30 km (b,d) SMF transmission.} 
   \label{fig:normal} 
\end{figure}

In the first experiments, we were demultiplexing standard data signals. The symbols were shaped digitally by raised-cosine filters with $1.0$ roll-off factor and baseband width of 4 GHz, for each low symbol rate channel. Thus, in the 24 GHz channel bandwidth, an aggregate data rate of 12 GBd was multiplexed with three optical sinc-sequences, each modulated with a 4 GBd data stream, corresponding to a 0.5 symbols/sec/Hz spectral efficiency. The result for the data content in one of the three sinc sequences can be seen in Fig. \ref{fig:normal} for QPSK and 16-QAM data formats transmitted via 10 and 30 km of fiber. The other signals can be demultiplexed by $120^\circ$ phase shifts of the RF signal driving the modulator. The eye diagrams correspond to the in-phase component.  The corresponding signal metrics such as $Q$-factors and average-EVM values have been presented in Table. \ref{table:1}. All the measurements of signal metrics were done following the decision threshold method \cite{Bergano1993} for $3\times 10^5$ recorded bits. For these experiments, a direct measurement of the bit-error-rate (BER) was not possible, since no errors occurred during the measurement. To avoid a long data acquisition time, the BER values can be estimated from the $Q$-factors (in linear scale) for two-level signals following the relationship $\mathrm{BER}_{est}\approx(1/2) \mathrm{erfc}(Q/\sqrt{2})$, \cite{SchmogrowEVM}. The measured $Q$-factors for the demultiplexed QPSK signals after 30 km transmission correspond to estimated BER values in the order of $10^{-17}$. These values are well below the $4.5 \times 10^{-3}$ limit for the hard-decision forward error correction (HD-FEC) codings with 7\% overhead \cite{Chang2010}. \par
\renewcommand{\arraystretch}{1.6}
\begin{table}[htp]
\centering
\caption{Measured performance metrics of one demultiplexed 4 GBd data channel from a 12 GBd multiplexed channel with a carrier frequency of 193.4 THz.}
 \begin{tabular}{| c | c | c | c |} 
 \hline
 Format & Distance & $Q$-factor & Average EVM \\ [-1.5ex]
 & (km) & (dB) & (\%) \\
\hline 
 \multirow{2}{*}{QPSK} & \multirow{2}{*}{10} &I $=20.42887\pm1.04338$ & \multirow{2}{*}{$9.9889\pm0.4216$}\\
\cline{3-3}
 & & Q $=19.96499\pm0.51241$& \\
 \hline
 \multirow{2}{*}{QPSK} & \multirow{2}{*}{30} &I $=18.46182\pm0.64159$ & \multirow{2}{*}{$13.1057\pm0.5449$}\\
\cline{3-3}
 & & Q $=18.268\pm0.25387 $& \\
 \hline
 \multirow{2}{*}{16-QAM} & \multirow{2}{*}{10} &I $=14.75606\pm0.40704$ & \multirow{2}{*}{$6.6485\pm0.1887$}\\
\cline{3-3}
 & & Q $=13.58524\pm0.26846 $& \\
 \hline
 \multirow{2}{*}{16-QAM} & \multirow{2}{*}{30} &I $=12.03721\pm0.22764$ & \multirow{2}{*}{$9.27\pm0.1891$}\\
\cline{3-3}
 & & Q $=11.102\pm0.16626 $& \\
 \hline
\end{tabular}
\label{table:1}
\end{table}

In another set of experiments, each low-capacity channel was modulated with Nyquist signals of $B/4=8$ GBd symbol rate having 4 GHz baseband width. Here the symbols were shaped by sinc-shaped pulses. We used digital filtering by raised-cosine filters with $0.0$ roll-off factor and baseband width of 4 GHz for this purpose. Again, three of such signals were multiplexed via optical sinc-sequences of 24 GHz bandwidth. Thereby, a 24 GBd Nyquist-QPSK channel was created with a theoretically maximum spectral efficiency of 1 symbol/sec/Hz. In the experiments we were restricted by the bandwidth of our arbitrary waveform generator to a maximum data rate of 48 Gbit/s. Thus, modulation formats higher than QPSK could not be generated in this bandwidth.  \par
\begin{figure}[!ht]
   \centering
   \includegraphics[width=0.85\columnwidth]{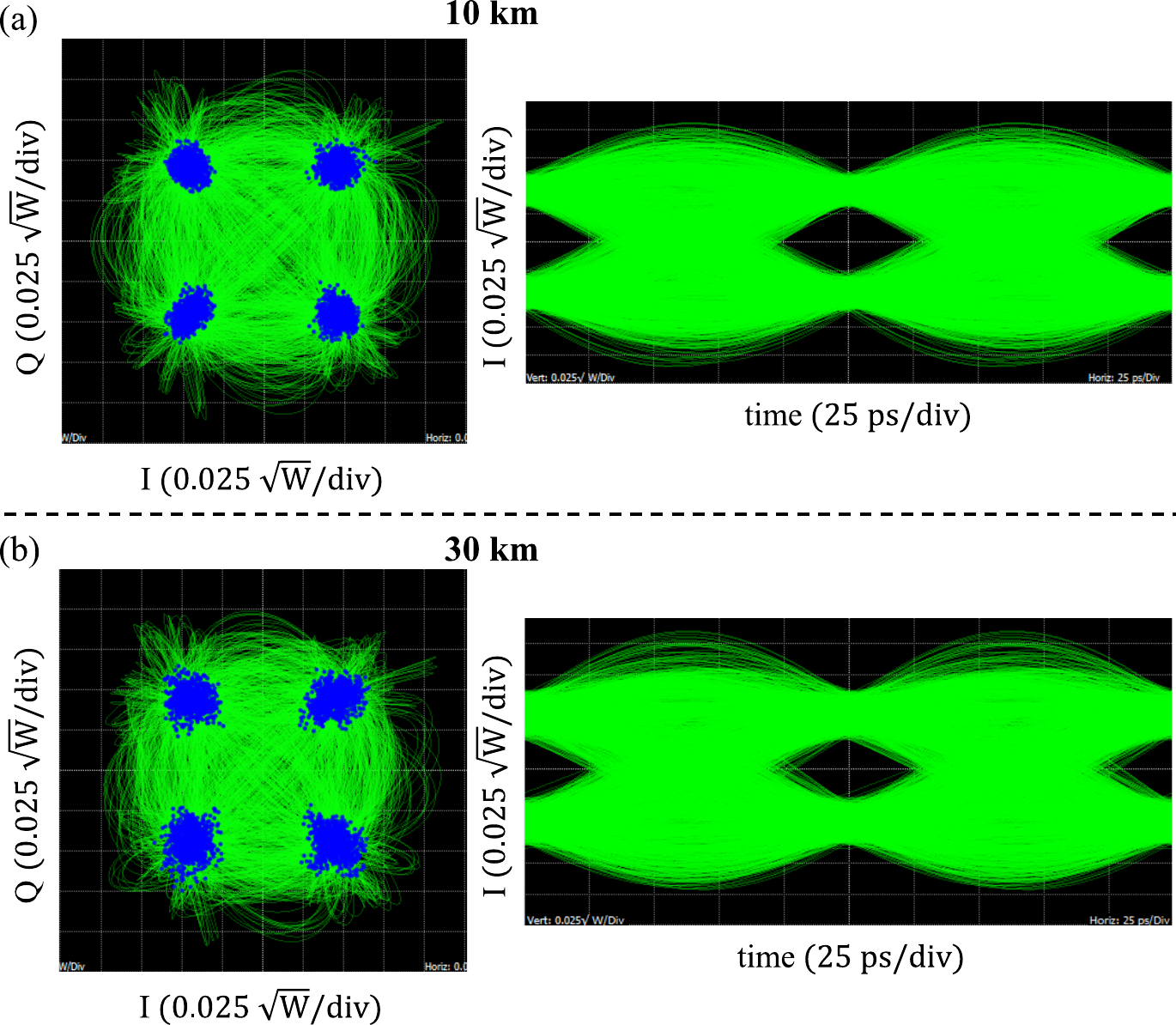}
   \caption{ Measured Symbol constellations and eye diagrams for one demultiplexed 8 GBd Nyquist-QPSK signal from a Nyquist channel of 24 GHz bandwidth around 193.4 THz after (a) 10 km and (b) 30 km SMF transmission.}
   \label{fig:sinc} 
\end{figure}
\begin{figure}[!ht]
   \centering
   \includegraphics[width=0.75\columnwidth]{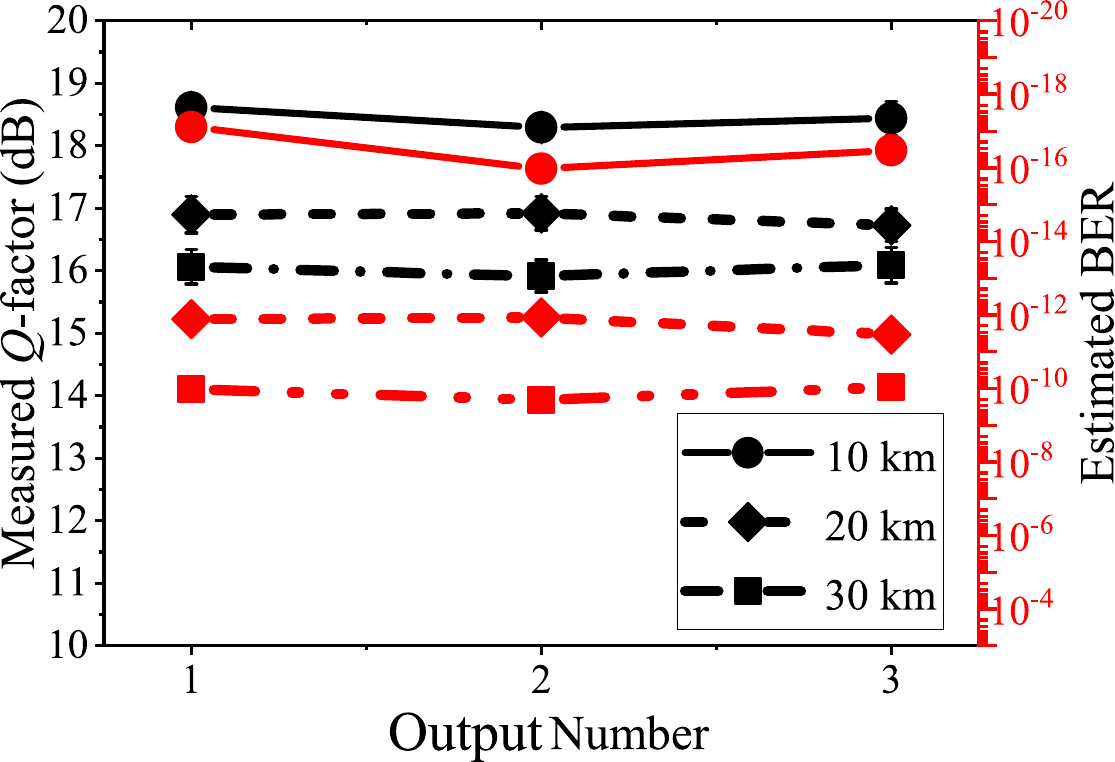}
   \caption{Measured $Q$-factors (in black) and estimated BER values (in red) for different transmission ranges of all three 8 GBd Nyquist-QPSK signals demultiplexed from the 24 GBd Nyquist channel.}
   \label{fig:Q} 
\end{figure}

For one demultiplexed Nyquist-QPSK signal branch with a symbol rate of 8 GBd, the symbol constellations and eye diagrams are presented in Fig. \ref{fig:sinc}, with the measured signal metrics presented in Table. \ref{table:2} after 10 km and 30 km of SMF transmission. The optical signal-to-noise ratio values before demultiplexing were around 37 dB and 33 dB respectively after 10 km and 30 km of fiber transmission for the measurements involving 8 GBd QPSK signals. As shown in Fig. \ref{fig:Q}, all the three demultiplexed 8 GBd Nyquist-QPSK signals from the 24 GBd Nyquist channel have similar performance for a certain distance of fiber transmissions. All of the measured BER values are far away from the limit for HD-FEC coding with 7 \% overhead. \par
\renewcommand{\arraystretch}{1.6}
\begin{table}[htp]
\centering
\caption{Measured performance metrics of one demultiplexed 8 GBd Nyquist-QPSK signal from the 24 GBd Nyquist channel around 193.4 THz.}
 \begin{tabular}{| c | c | c | c |} 
 \hline
 Format & Distance & $Q$-factor & Average EVM \\ [-1.5ex]
 & (km) & (dB) & (\%) \\
\hline 
\multirow{3}{*}{Nyquist-} & \multirow{2}{*}{10} &I $=18.61079\pm0.17767$ & \multirow{2}{*}{$12.3597\pm0.1567$}\\
\cline{3-3}
  & & Q $=18.29439\pm0.12573$& \\
 \cline{2-4}
  \multirow{1}{*}{QPSK}& \multirow{2}{*}{30} &I $=16.22659\pm0.36605$ & \multirow{2}{*}{$15.9707\pm0.9316$}\\
\cline{3-3}
 & & Q $=16.1480\pm0.3467$& \\
 \hline
\end{tabular}
\label{table:2}
\end{table}

\begin{figure}[!ht]
   \centering
   \includegraphics[width=\columnwidth]{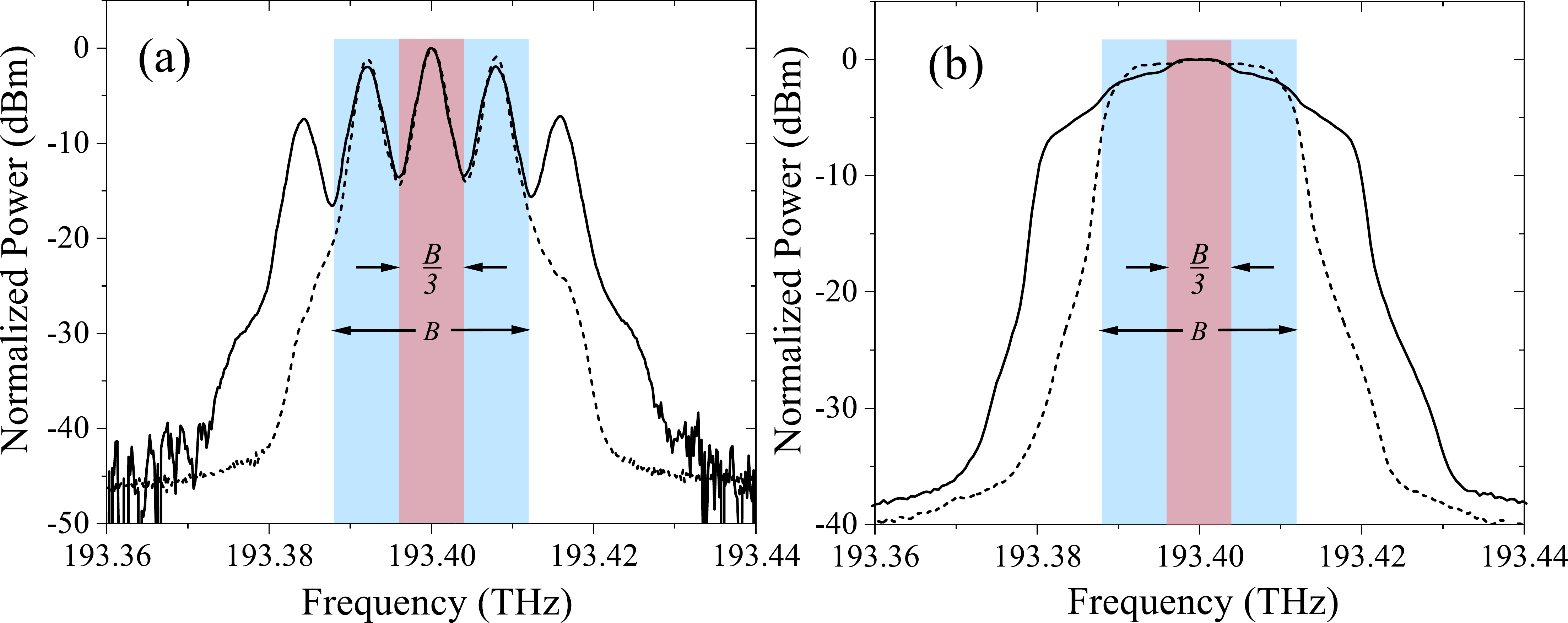}
   \caption{Measured Optical spectra before (dashed curves) and after (solid curves) demultiplexing without any transmission fiber of significant length. For (a) three 4 GBd QPSK signals were orthogonally time multiplexed in a 24 GHz bandwidth (blue rectangle). For (b) three 8 GBd Nyquist-QPSK signals were multiplexed in the same bandwidth. The red rectangle shows the spectral region of the data information after demultiplexing.}
   \label{fig:spec1} 
\end{figure}

The optical spectra of the multiplexed (dashed lines) and demultiplexed (solid lines) signals around a single carrier at 193.4 THz for demultiplexing with a flat, three-line rectangular comb can be seen in Fig. \ref{fig:spec1}. Please note, that the demultiplexing is based on a convolution with the three-line comb. Thus, the spectrum after multiplexing has a rectangular shape (dashed lines) and the spectral content broadens due to the demultiplexing . However, the whole information of the signal is just in the $B/3$ fraction of the bandwidth around the carrier (red rectangle), resulting in a required electrical detector bandwidth of $B/6$. For Fig. \ref{fig:spec1}(a), three 4 GBd QPSK signals were multiplexed within the 24 GHz bandwidth and for Fig. \ref{fig:spec1}(b), three Nyquist-QPSK signals of 8 GBd symbol rate were multiplexed in the same bandwidth (dashed curve). \par

\begin{figure}[!ht]
   \centering
   \includegraphics[width=\columnwidth]{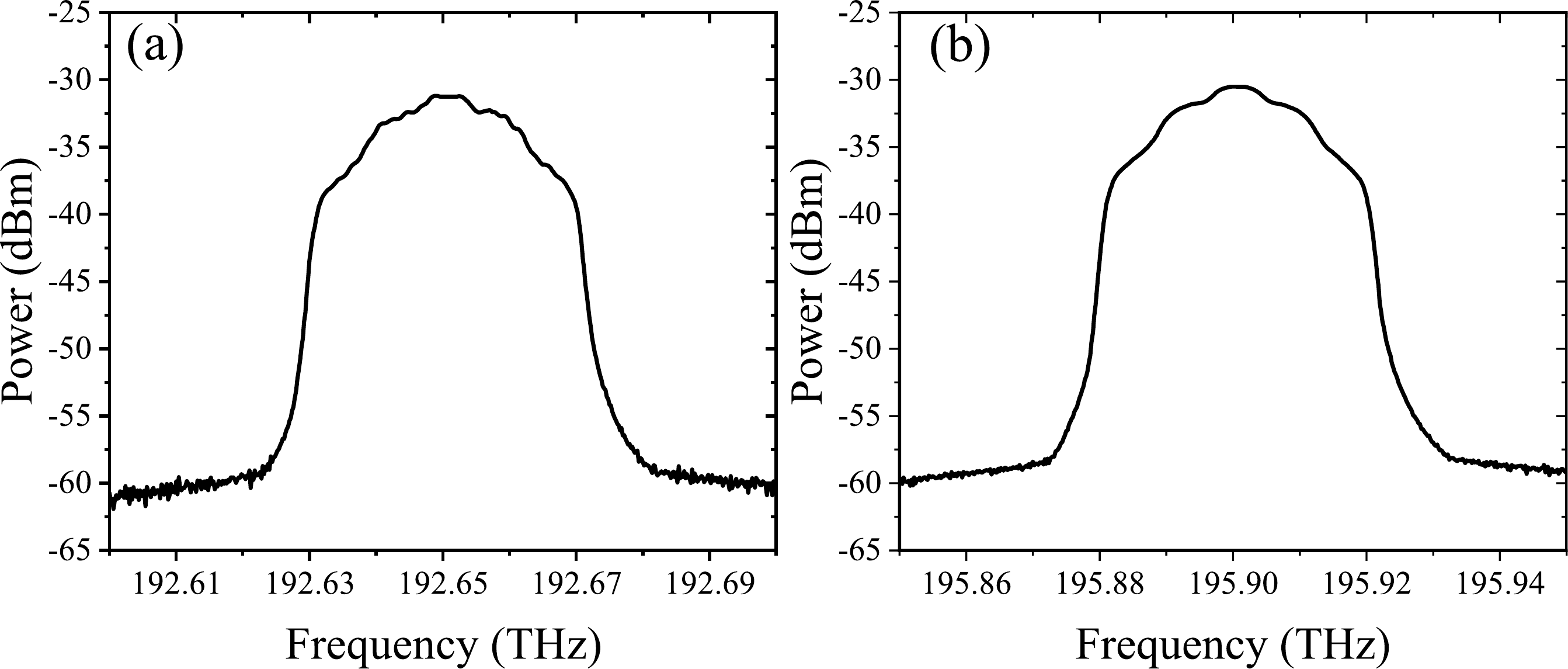}
   \caption{Measured Optical spectra at central frequencies of (a) 192.65 THz, and (b) 195.90 THz after demultiplexing a 24 GBd Nyquist-QPSK channel to three 8 GBd Nyquist QPSK channels for 30 km SMF transmission.}
   \label{fig:spec2} 
\end{figure}

Furthermore, to evaluate the optical operation bandwidth, Nyquist-QPSK signal transmission experiments were carried out at different carrier frequencies. We chose three nominal central frequencies of 195.90 THz (1530.3341 nm), 193.40 THz (1550.116 nm), and 192.65 THz (1556.151 nm) in accordance with the recommendation of the international telecommunication union (ITU-T G.694.1(10/20)). The experimental setup adopted here is the same as the previous one except for the operating laser frequency. The spectrum at 193.4 THz (1550.116 nm) has already been presented above. The measured optical spectra for the other two frequencies after demultiplexing have been presented in Fig. \ref{fig:spec2}, and the data transmission metrics are summarized in  Table. \ref{table:3} for 30 km SMF transmission.\par
It can be seen that the performance of the demultiplexer is very similar at this wide frequency range spanning almost the full C-band (1530 nm - 1560 nm) with estimated BER values of the order of $10^{-10}$ after 30 km of SMF transmission. A limitation in the operational wavelength range would mainly be determined by wavelength-dependent grating couplers for optical coupling to the chip and on-chip multimode interferometers used for splitting and combining the MZM arms. However, the basic principle can be implemented for any carrier frequency of interest belonging to any communication band. \par
\begin{figure}[!ht]
   \centering
   \includegraphics[width=\columnwidth]{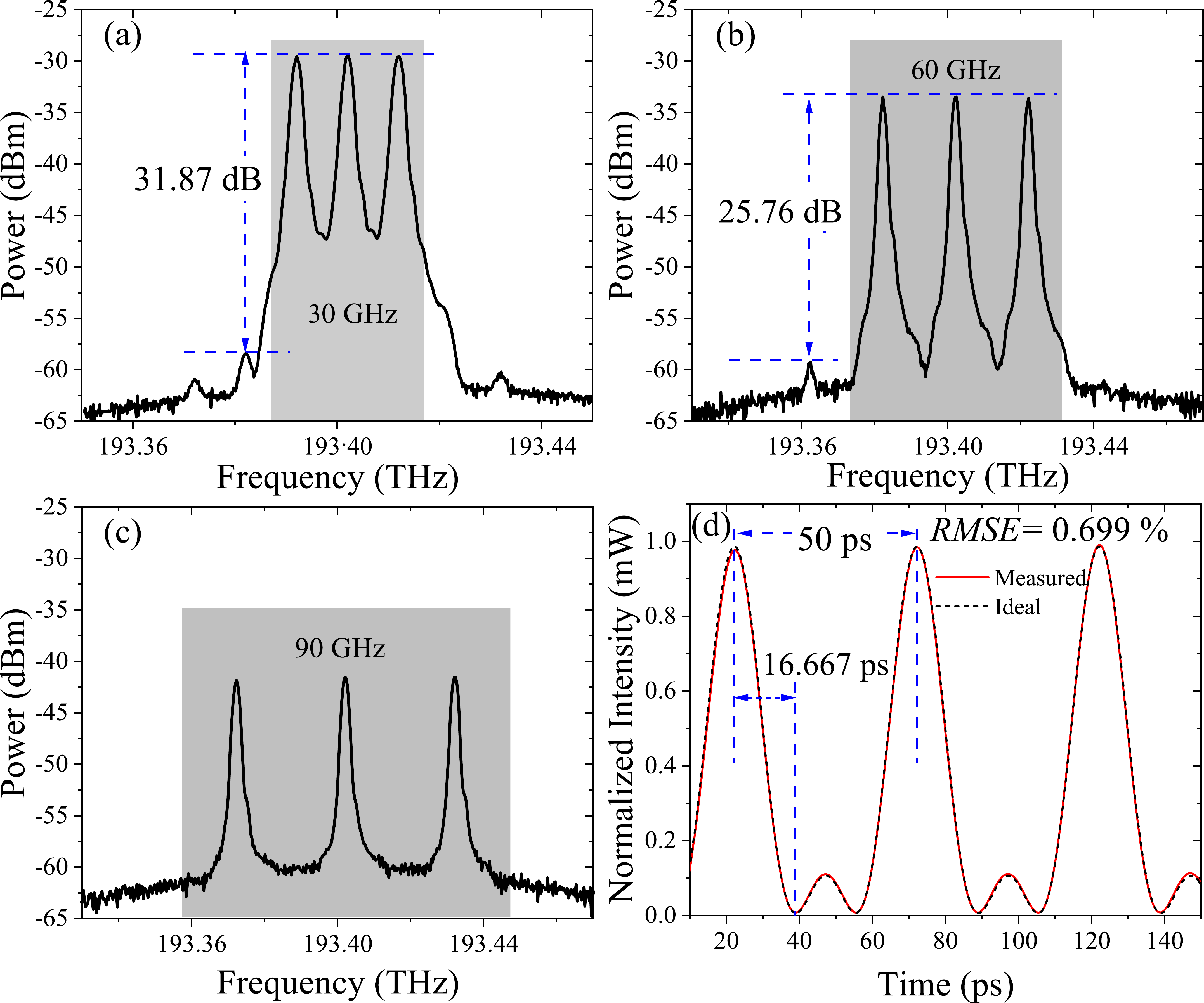}
   \caption{Measured Optical spectra of non-filtered three-line rectangular phase locked OFCs of (a) 10 GHz, (b) 20 GHz, and (c) 30 GHz spacing under identical RF input power. (d) Measured Nyquist pulse sequence of 60 GHz bandwidth, generated with the 16 GHz modulator. The values for the power difference between the comb and next unwanted sideband are noted.}
   \label{fig:combs} 
\end{figure}

\renewcommand{\arraystretch}{1.6}
\begin{table}[htp]
\centering
\caption{Signal metrics for the demultiplexed 8 GBd Nyquist QPSK signal at different
central frequencies after 30 km SMF transmission}
 \begin{tabular}{| c | c | c |} 
 \hline
 Central frequency & $Q$-factor & Average EVM \\ [-1.5ex]
 (THz) & (dB) & (\%) \\
\hline 
\multirow{2}{*}{192.65} &I $=15.92985\pm0.29706$ & \multirow{2}{*}{$16.1480\pm0.3467$}\\
\cline{2-2}
  & Q $=15.91293\pm0.16985$& \\
 \hline
 \multirow{2}{*}{193.4} &I $=16.22659\pm0.36605$ & \multirow{2}{*}{$15.9707\pm0.9316$}\\
\cline{2-2}
  & Q $=16.1480\pm0.3467$& \\
 \hline
\multirow{2}{*}{195.95} &I $=16.174\pm0.43588$ & \multirow{2}{*}{$16.0940\pm0.2871$}\\
\cline{2-2}
  & Q $=16.13527\pm0.50083 $& \\
 \hline
\end{tabular}
\label{table:3}
\end{table}
The demultiplexing of the Nyquist data channel is based on the convolution between the incoming optical signal with the bandwidth $B$ and the three-line frequency comb, generated in the modulator by driving it with one single RF frequency $\Delta f$. For a CW optical line as input, a flat three-line frequency comb with a frequency spacing of $\Delta f$ would be the result of the convolution. Thus, as long as the modulator can generate this three-line comb for a CW input, it would be able to demultiplex a corresponding Nyquist data signal with a bandwidth of $3\Delta f$. Therefore, to demonstrate the tunability of the pulse repetition rate and bandwidth, we have generated flat rectangular three-line optical frequency combs (OFCs) of different frequency spacings $\Delta f$. The bandwidth of the comb ($3\Delta f$) defines the maximum possible aggregate symbol rate per carrier for demultiplexing. The input radio frequency to the modulator was varied for an unmodulated CW optical input, and waveform measurements were carried out with an oscilloscope after photodetection. The output OFCs of bandwidth ($B=3\Delta f$) are presented in Fig. \ref{fig:combs}(a)-\ref{fig:combs}(c) for input radio frequencies of $\Delta f =$10 GHz, 20 GHz, and 30 GHz respectively. The flatness of the measured combs was around 0.1 dB. Although the modulator's 3 dB electro-optic bandwidth is 16 GHz (see Fig. \ref{fig:chip}(d)), OFCs with a much higher spacing could be generated. The RF power to the modulator was kept constant at 4 dBm while the input voltage to the thermal phase shifters was varied to suppress the carrier relative to the sideband in order to get the desired flat comb. However, this over-driving comes at the expense of decreased optical output power of the comb as evident from Fig. \ref{fig:combs}. However, even for the 90 GHz comb the signal to noise ratio is still around 20 dB, which should be sufficient for demultiplexing\par 
Due to nonlinearities associated with the carrier dispersion effect and nonlinear transfer function, silicon PN junction modulators have a higher nonlinear response compared to LiNbO\textsubscript{3} modulators based on the linear electro-optic effect (Pockels effect). These nonlinearities result in higher order sidebands in the modulated spectra. The difference between the comb power and the next unwanted sideband is noted for Fig. \ref{fig:combs}(a) and \ref{fig:combs}(b). It is worth noting that no optical filters were used to suppress the unwanted sidebands and the suppression was very high. The results show considerable improvements upon the reported ones for silicon modulators \cite{Liu2020b,Liu2021}. In Fig. \ref{fig:combs}(d) a measured pulse waveform is presented for an OFC of 20 GHz spacing, resulting in a 60 GHz comb as shown in Fig. \ref{fig:combs}(b). It resembles the ideal pulse waveform (in black-dashed) very closely with a root mean square error (RMSE) of 0.699\%. Pulses beyond this bandwidth were not measurable with available experimental resources.
\section{Conclusions}
In conclusion, we have experimentally demonstrated real-time reconfigurable demultiplexing of quadrature amplitude modulated Nyquist channels in silicon photonics over the entire C-band, where all the parameters like channel selection and channel bandwidth can be electrically controlled. Due to limitations in the transmitter electronics we were restricted to single carrier line rates of 48 Gbit/s. However, with a modulator having 16 GHz 3 dB bandwidth, it was possible to generate three-line rectangular phase locked frequency combs with 30 GHz comb spacing leading to 90 GHz sinc-pulse sequences. It is the highest bandwidth flat optical three-line frequency comb reported so far in silicon photonics. The technology platform facilitates monolithic integration of photonic and electronic components on the same silicon substrate. Among other devices, this technology offers silicon germanium hetero-bipolar transistors with up to 220 GHz transit frequency and germanium photodiodes with more than 60 GHz electro-optical 3 dB bandwidth \cite{IHP}. Hence, the presented Si-MZM based demultiplexing technique enables the demultiplexing of Nyquist signals with single carrier symbol rates up to 90 GBd (corresponding to five times the electro-optic bandwidth of the modulator) in a coherent Nyquist transceiver with three parallel silicon photonic receivers of only 15 GHz detection bandwidths.

\ifCLASSOPTIONcaptionsoff
  \newpage
\fi

\bibliography{report}
\bibliographystyle{MyIEEEtran}

\end{document}